\documentclass[twocolumn]{aastex61}
\usepackage{natbib}

\accepted{August 17, 2017}
\submitjournal{AJ}

\shorttitle{36\,GHz class I methanol maser emission towards NGC\,4945}
\shortauthors{McCarthy et al.}

\begin{document}
	
	\title{Detection of 36\,GHz class I methanol maser emission towards NGC\,4945}
	
	\correspondingauthor{Tiege McCarthy}
	\email{tiegem@utas.edu.au}
	
	\author[0000-0001-9525-7981]{Tiege P McCarthy}
	\affil{School of Physical Sciences, University of Tasmania, Hobart, Tasmania 7001, Australia}
	\affil{Australia Telescope National Facility, CSIRO, PO Box 76, Epping, NSW 1710, Australia}
	
	\author[0000-0002-1363-5457]{Simon P. Ellingsen}
	\affil{School of Physical Sciences, University of Tasmania, Hobart, Tasmania 7001, Australia}
	
	\author[0000-0002-5435-925X]{Xi Chen}
	\affiliation{Shanghai Astronomical Observatory, Chinese Academic of Sciences, Shanghai 200030, China}
	\affiliation{Center for Astrophysics, GuangZhou University, Guangzhou 510006, China}

	\author[0000-0002-4047-0002]{Shari L. Breen}
	\affiliation{Sydney Institute for Astronomy (SIfA), School of Physics, University of Sydney, NSW 2006, Australia}
	
	\author[0000-0002-4047-0002]{Maxim A. Voronkov}
	\affiliation{Australia Telescope National Facility, CSIRO, PO Box 76, Epping, NSW 1710, Australia}	
	
	\author[0000-0003-0196-4701]{Hai-hua Qiao}
	\affiliation{National Time Service Center, Chinese Academy of Sciences, Xi'An, Shaanxi, 710600, China}
	\affiliation{Shanghai Astronomical Observatory, Chinese Academic of Sciences, Shanghai 200030, China}

\begin{abstract}
\noindent We have searched for emission from the 36.2\,GHz ($4_{-1} \rightarrow 3_0$E) methanol transition towards NGC\,4945, using the Australia Telescope Compact Array. 36.2\,GHz methanol emission was detected offset south-east from the Galactic nucleus. The methanol emission is narrow, with a linewidth $<$10\,km\,s$^{-1}$, and a luminosity five orders of magnitude higher than Galactic class I masers from the same transition. These characteristics combined the with physical separation from the strong central thermal emission suggests that the methanol emission is a maser. This emission is a factor of $\sim90$ more luminous than the widespread emission detected from the Milky Way central molecular zone (CMZ). This is the fourth detection of extragalactic class I emission, and the third detection of extragalactic 36.2\,GHz maser emission. These extragalactic class I methanol masers do not appear to be simply highly luminous variants of Galactic class I emission, and instead appear to trace large-scale regions of low-velocity shocks in molecular gas, which may precede, or be associated with, the early stages of large-scale star formation.

\end{abstract}
\keywords{masers -- radio lines: ISM -- galaxies: starburst -- galaxies: individual (NGC4945)}

\section{Introduction} \label{sec:intro}

\begin{table*}
	\begin{center}
		\caption{36.2\,GHz methanol and continuum emission toward NGC\,4945. Numbers in parenthesis indicate 3$\sigma$ error in the value and locations of emission are accurate to $<0.4$ arcseconds.}
		\begin{tabular}{ccccccc} \hline
			\multicolumn{1}{c}{\bf} & \multicolumn{1}{c}{\bf RA (J2000)}  & \multicolumn{1}{c}{\bf Dec (J2000)} & \multicolumn{1}{c}{\bf Peak Flux} & \multicolumn{1}{c}{\bf Integrated Flux} & \multicolumn{1}{c}{\bf Peak Velocity} & \multicolumn{1}{c}{\bf Velocity Range}  \\
			& \multicolumn{1}{c}{\bf $h$~~~$m$~~~$s$}& \multicolumn{1}{c}{\bf $^\circ$~~~$\prime$~~~$\prime\prime$} & (mJy) & (mJy\,km\,s$^{-1}$) & (km\,s$^{-1}$) & (km\,s$^{-1}$) \\   \hline
			Methanol & 13 05 28.093 & $-$49 28 12.306 & 43.8 & $256\pm35$  & 674 & 660 -- 710 \\ 
			Continuum & 13 05 27.467 & $-$49 28 04.797 & 349 &  $385\pm40$  &  & \\   \hline
		\end{tabular} \label{tab:emission}		
	\end{center}
\end{table*}

Methanol maser emission is a commonly observed phenomenon in the Milky Way. The species has a rich spectrum of masing transitions, allowing it to be utilised as a powerful tool for probing high-mass star-formation regions in our Galaxy. Galactic methanol masers have now been detected in more than twelve hundred sources \citep[e.g][]{Ellingsen+05,Caswell+10,Caswell+11, Voronkov+14, Breen+15, Green+10, Green+12a, Green+17}. The masing transitions of methanol are empirically divided into two classes, based on the pumping mechanism responsible for excitation, collisional for class I and radiative for class II. Galactic class I masers are typically associated with the interaction of shocked gas with molecular clouds, driven by outflows or expanding H \small{II} regions \citep{Kurtz+04, Cyganowski+09, Cyganowski+12, Voronkov+10a, Voronkov+14}. These shocks aid the release of methanol into the gas phase, along with warming and compressing the gas, which produces optimal masing conditions \citep{Voronkov+10a}. Class I masers are often observed distributed across many different sites within one star-formation region, on scales up to 1\,pc \citep{Voronkov+14}. Galactic class II masers are closely associated with young stellar objects (YSOs), which are often accompanied by OH and H$_2$O masers, and are usually observed in one or two compact sites within a star-formation region \citep[e.g][]{Ellingsen06, Caswell+10, Breen+10b}. Class II masers are exclusively observed towards high-mass star-formation regions \citep{Breen+13b}, while class I masers have been observed associated towards low- and high-mass star-formation regions as well as supernova remnants \citep[e.g.][]{Kalenskii+10,Pihlstrom+14}.

Despite the abundance of masing methanol observed within our Galaxy, extragalactic methanol masers are a rarely observed phenomenon. Class II maser emission (either 6.7 or 12.2\,GHz) has been observed in the Large Magellanic Cloud and M31, with what appear to be similar properties to their Galactic counterparts  \citep{Green+08, Ellingsen+10, Sjouwerman+10}. Extragalactic class I emission has been previously observed at 36.2\,GHz in NGC\,253 and Arp\,220 \citep{Ellingsen+14, Chen+15}, and in NGC\,1068 at 84.5\,GHz \citep{Wang+14}. Two of these reported extragalactic class I sources (Arp\,220 and NGC\,1068) are luminous enough to be considered megamasers ($> 10\,\text{L}_{\sun}$). In contrast to the extragalactic class II masers, these class I sources do not appear to be the result of similar processes as their Galactic equivalents, instead they appear to trace regions of large-scale molecular in-fall \citep{Ellingsen+submitted}.

The 36.2\,GHz $4_{-1} \rightarrow 3_0 $E methanol transition is one of the most common class I methanol masers observed towards star-formation regions in the Milky Way \citep{Haschick+89b, Voronkov+14}. Widespread emission from this same transition has also been observed in the inner region of our Galaxy, with over 350 independent sites detected within a $160 \times 43$\,pc region resulting in an integrated luminosity $>5600$\,Jy\,km\,s$^{-1}$ \citep{Yusef-Zadeh+13}. Photo-desorption of methanol from cold dust by cosmic rays has been suggested as the mechanism producing the abundance of methanol required to produce these levels of observed methanol maser emission \citep{Yusef-Zadeh+13}. It is reasonable to expect that a similar phenomenon may exist in other other galaxies, with \citet{Ellingsen+14} suggesting that galaxies with enhanced star-formation may display this emission over even larger volumes, making them more readily detectable.

Here we report a search for the class I 36.2\,GHz and class II 37.7\,GHz transition towards the central region of NGC\,4945. This nearby \citep[assumed distance of 3.7$\pm0.3$\,Mpc;][]{Tully+13}  galaxy is classified as a barred spiral and starburst type. In addition to being classified as starburst, the nucleus of NGC\,4945 is also classified as an AGN, and is the strongest source of hard x-rays observed from Earth \citep{Done&96}. Analysis of near- and far-infrared observations indicates that the AGN is heavily obscured, and that the starburst process is the primary source of energy for exciting photo-ionized gas \citep{Marconi+00, Spoon+00, Spoon+03, Perez-Beauputis+11}. NGC\,4945 has a star-formation rate of $4.35\pm0.25$ M$_{\sun}$ yr$^{−1}$ \citep{Bendo+16}, almost a factor of 3 more than the $1.65\pm0.19$ M$_{\sun}$ yr$^{−1}$  observed in the Milky Way \citep{Licquia+15}.

\section{Observations} \label{sec:obs}

\begin{figure}
	\includegraphics[width=\linewidth]{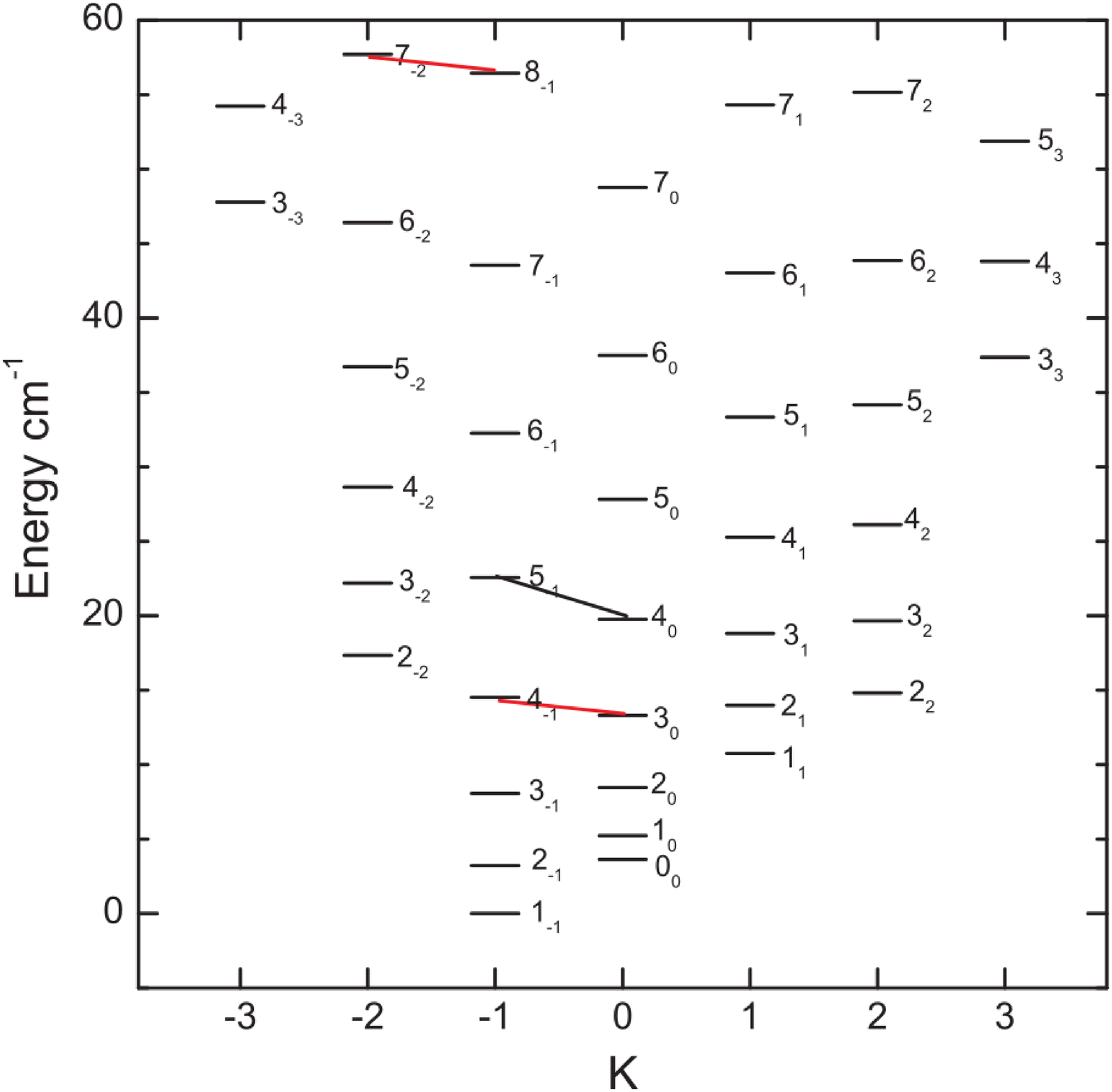}
	\caption{Rotational energy level diagram for E-type methanol originally presented in \citet{Chen+15} and based on the CDMS catalogue \citep{Muller+01}. The three marked transitions correspond to the transitions of class I methanol observed masing in extragalactic sources \citep{Ellingsen+14, Wang+14, Chen+15}. The red marked transitions are those searched for towards NGC\,4945, NGC\,253 and Arp\,220, with the black marked transition representing the 84.5\,GHz transition detected in NGC\,1068 \citep{Wang+14}.}
	\label{fig:elevel_diag}
\end{figure}

The observations were made using the Australia Telescope Compact Array (ATCA) on 2015 August 25 and 26 (project code C2879). The observations utilised the EW352 array, with minimum and maximum baselines of 31 and 352 metres resulting in a synthesised beam of approximately $26\times4$ arcseconds at 36.2\,GHz (elongation caused by limited hour-angle coverage). The Compact Array Broadband Backend \citep{Wilson+11} was configured in hybrid mode CFB 1M/64M, with both 2.048\,GHz bands centred on 36.85\,GHz. Two zoom bands were configured in the 64 MHz resolution IF, covering the rest frequencies of the $4_{−1} \rightarrow 3_0 $E and $7_{−2} \rightarrow 8_{−1} $E transitions of methanol (marked in Figure \ref{fig:elevel_diag}), for which we adopted rest frequencies of 36.169265 and 37.703700\,GHz, respectively \citep{Muller+04}. This resulted in a velocity range of -350 to 1200 km\,s$^{-1}$ (barycentric) with a spectral resolutions of 0.26 km\,s$^{-1}$ at 36.2\,GHz. We have used the barycentric reference frame for all velocities reported in this paper. The FWHM of the primary beam of the ATCA antennas at 36.2\,GHz is approximately 70 arcseconds and for an assumed distance of 3.7\,Mpc to NGC\,4945 this corresponds to a linear scale of 1200\,pc.  Our observations used a single pointing centred on the nucleus of the galaxy and so are only sensitive to methanol maser emission within 600\,pc of the centre of the galaxy.

The data were reduced with {\sc miriad} using the standard techniques for ATCA 7mm spectral line observations. Amplitude calibration was with respect to PKS B1934-648, and PKS B1253-055 was observed as the bandpass calibrator. The data were corrected for atmospheric
opacity and the absolute flux density calibration is estimated to be accurate to better than 30\%. The observing strategy interleaved ten minutes onsource with two minute observations of the phase calibrator (J1326-5256). The data were self-calibrated (phase and amplitude) using the continuum emission from the central region of NGC\,4945.  After self-calibration we used continuum subtraction (modelled using the spectral channels without maser emission) to isolate the spectral line and continuum emission components. The total duration on-source for NGC\,4945 was 1.68 hours. Molecular line emission in NGC\,4945 is observed between approximately 300 and 800\,km\,s$^{-1}$ \citep{Ott+01} and the velocity range of our imaging covered the barycentric velocities between 200 and 1000\,km\,s$^{-1}$, with a spectral resolution of 1\,km\,s$^{-1}$, and average RMS noise of $\sim$2.2 mJy beam$^{-1}$ in each spectral channel. Positions were determined using the {\tt MIRIAD} task imfit, which reports the peak value and location of a two-dimensional Gaussian fit for the emission in a given velocity plane within the spectral line cube. 

\section{Results} \label{sec:results}

We detected emission from the 36.2\,GHz methanol transition as well as 7 mm continuum emission towards NGC\,4945 (see Table \ref{tab:emission}). The 37.7\,GHz methanol transition was not detected from this source. We also looked for thermal emission from molecular lines within the bandpass of the recorded data and detected weak thermal emission from the HC$_3$N (J = 4 -- 3) transition.

\begin{figure*}
	\begin{minipage}[h]{0.50\linewidth}
		\centering
		\includegraphics[scale=0.38]{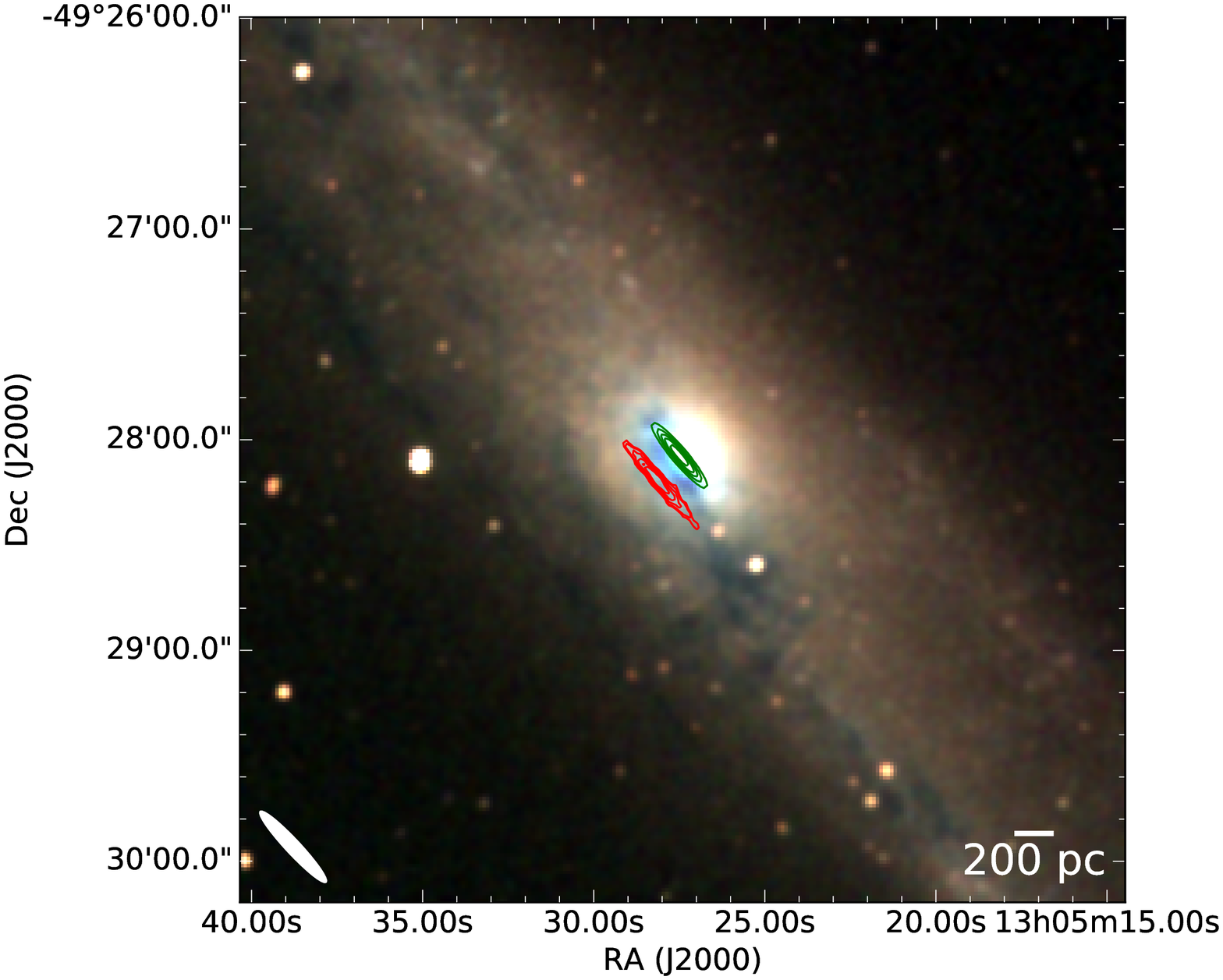}
	\end{minipage}
	\begin{minipage}[h]{0.50\linewidth}
		\centering
		\includegraphics[scale=0.75]{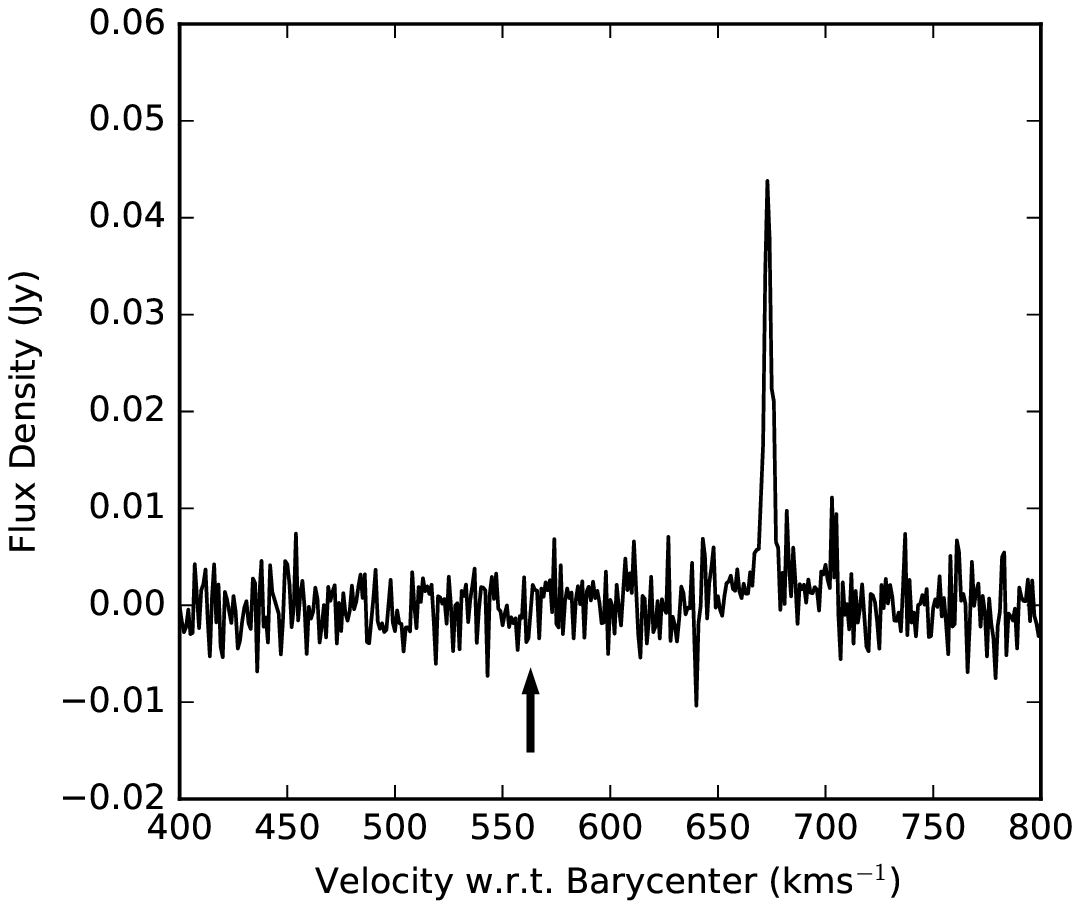}
	\end{minipage}
	\caption{Left: 36.2\,GHz emission methanol emission (red contours 45\%, 50\%, 60\%, 70\%, and 80\% of the 256 mJy\,km\,s$^{-1}$ beam$^{-1}$) and the 7mm continuum emission (green contours 15\%, 30\%, 45\%, 60\%, and 80\% of the 385 mJy\,km\,s$^{-1}$ beam$^{-1}$) with background 2MASS 3 colour image(J-,H- and K-band in red, green and blue respectively). The white ellipse describes the synthesised beam size for our observations. Right: 36.2\,GHz spectrum from the region of peak emission within our spectral line cube (imaged at 1\,km\,s$^{-1}$). Vertical arrow indicates the systemic velocity of NGC\,4945.}
	\label{fig:plot&spec}
\end{figure*}

The 7mm continuum emission is observed at the location of the galactic nucleus of NGC\,4945 \citep[location of the H$_2$O megamaser;][]{Greenhill+97}, while the 36.2\,GHz methanol emission is observed to be offset by $\sim10$ arcseconds (see Figure \ref{fig:plot&spec}), perpendicular from the position angle of the galactic disk \citep[$43^{\text{o}}\pm2^{\text{o}}$;][]{Dahlem+93}. We can more reliably measure offsets perpendicular to the disk, as the major-axis of our elongated synthesised beam is parallel to the position angle of the galaxy. This angular offset corresponds to a linear projected separation of $174\pm14$\,pc at an assumed distance of 3.7$\pm0.3$\,Mpc \citep{Tully+13}. The east-west array configuration combined with limited hour-angle range has made determination of an exact position for the methanol emission difficult due to a highly elongated synthesised beam. However, as the emission appears to compact, the position reported by imfit provides a suitable estimate until follow-up observations can be made with more complete uv coverage.


In contrast to the reasonably strong detection in the 36.2\,GHz zoom, the 37.7\,GHz has no detected emission to a level of around 2.5 mJy beam$^{-1}$. Similar to \citet{Ellingsen+14}, the 37.7\,GHz transition was only included as it could be simultaneously observed with the 36.2\,GHz emission and was not expected to be detected due to its rarity in Galactic sources.

\section{Discussion} \label{sec:disc}

\subsection{Masing methanol in NGC\,4945}

This is the third time the 36.2\,GHz methanol transition ($4_{-1} \rightarrow 3_0$E) has been observed in an extragalactic source \citep{Ellingsen+14, Chen+15}. In the two previous cases the emission has been convincingly described as resulting from maser processes. It is important for us to justify that the 36.2\,GHz methanol emission observed towards NGC\,4945 is indeed maser emission, rather than the result of thermal processes. Thermal methanol has not previously been observed towards NGC\,4945.

Narrow linewidths are a prominent characteristic of class~I methanol emission from within our Galaxy. The extragalactic 36.2\,GHz emission observed from NGC\,253 shares this property \citep[$\sim$10\,km\,s$^{-1}$;][]{Ellingsen+14}, however, the megamaser emission towards Arp\,220 is significantly more broad \citep{Chen+15}. For NGC\,4945 the 36.2\,GHz methanol emission is spread over a velocity range of 50\,km\,s$^{-1}$ (from 660 to 710\,km\,s$^{-1}$), dominated by a narrow (FWHM linewidth of $\sim8$\,km\,s$^{-1}$) peak component at 674\,km\,s$^{-1}$. There also appears to be a secondary component redshifted by approximately 30\,km\,s$^{-1}$ that is much weaker (factor of 4 dimmer), however, it shares a similar linewidth. Therefore, we observe similar emission characteristics in NGC\,4945 as displayed by the 36.2\,GHz emission in NGC\,253 \citep{Ellingsen+14}. In addition to the narrow linewidth, we observe no thermal emission associated with the location of the methanol emission, with the relatively weak thermal HC$_3$N emission (J = 4 -- 3) observed only towards the galactic nucleus. The velocity of this HC$_3$N emission covers the same range that molecular line emission is observed from towards NGC\,4945 \citep{Ott+01} In contrast to the HC3N (J=4-3) observations towards NGC\,253 \citep{Ellingsen+17}, in NGC4945 there does not appear to be a close association with the 36.2 GHz methanol emission.

Considering the integrated intensity observed for the methanol emission of 0.258 Jy\,km\,s$^{-1}$ , we determine an isotropic luminosity of 4.44 $\times$ 10$^7$ Jy\,km\,s$^{-1}$\,kpc$^2$. Compared with the 9.0 $\times$ 10$^7$ and 9.5 $\times$ 10$^{11}$ Jy\,km\,s$^{-1}$\,kpc$^2$ observed towards NGC\,253 and Arp\,220 respectively \citep[note, there is a typographical error in Table 2 of \citet{Chen+15} and that the units for the integrated intensity are mJy\,km\,s$^{-1}$, not Jy\,km\,s$^{-1}$;][]{Ellingsen+14, Chen+15}. It should be noted that all luminosity values reported here include the factor of $4\pi$, this was excluded from the value reported in \citet{Ellingsen+14}. \citet{Voronkov+14} presents data for a large number of Galactic 36.2\,GHz methanol masers and utilising the data presented in their table~3, along with distance estimates from \citet{Green+McClure11} we find 500~Jy\,km$\,s^{-1}$ kpc$^2$ to be a typical isotropic luminosity for this transition in Galactic high-mass star formation regions. If we consider 500\,Jy\,km\,s$^{-1}$ kpc$^2$ representative of Galactic class~I methanol masers, the 36.2\,GHz transition in NGC\,4945 is approximately five orders of magnitude more luminous. This luminosity corresponds to an integrated intensity of $6.1$ $\times$ 10$^5$ Jy\,km\,s$^{-1}$ at a distance of 8.5\,kpc, therefore, this methanol emission in NGC\,4945 is a factor of $\sim90$ greater than the emission in the Milky Way CMZ \citep{Yusef-Zadeh+13}.

The narrow FWHM linewidth combined with an isotropic luminosity comparable to that observed in NGC\,253, and angular separation from any detected strong thermal emission provides a strong indication that the 36.2\,GHz emission towards NGC\,4945 is a maser. Future observations of NGC\,4945 with better uv coverage will allow for more accurate determination of the size 36.2\,GHz emission region. This more accurate imaging, combined with observations of thermal methanol in the central region of NGC\,4945 will allow verification that this methanol emission is due to a maser process. 

\subsection{Properties of methanol maser environment}

H$_2$O megamasers have been extensively observed towards NGC\,4945 (in the 22 , 183 and 321\,GHz transitions), these megamasers are distributed within a circumnuclear accretion disk ($<1$\,pc from nucleus) oriented parallel to the position angle of the galactic disk \citep{Greenhill+97, Pesce+16, Hagiwara+16, Humphreys+16}. In contrast, the methanol maser emission is significantly offset to the south-east of the galactic nucleus. This offset from the galactic centre is consistent with the observed class I maser emission in Arp\,220 and NGC\,253 \citep{Ellingsen+14, Chen+15}. \citet{Ellingsen+14} observe offsets of 180 and 300\,pc for the north-east and south-west components respectively. Although the H$_2$O maser emission observed in NGC\,4945 is not physically associated with the 36.2\,GHz maser emission, the redshifted components share an overlapping velocity range \citep{Greenhill+97, Pesce+16, Hagiwara+16, Humphreys+16}.

Detailed investigations of the molecular clouds in NGC\,4945 have revealed an elliptical rotating molecular cloud complex extending approximately 400\,pc with major axis aligned with the galactic disk \citep{Cunningham&05, Chou+07}. The 36.2\,GHz maser emission lies within the outer edge region of the CO and HCN circumnuclear molecular clouds, and there is no particular enhancement in the quiescent, or dense gas molecular tracers at the location of the class I methanol maser emission \citep{Cunningham&05, Chou+07}. The 36.2\,GHz maser emission is covered by the 540 - 750\,km\,s$^{-1}$ velocity range of the $^{12}$CO(2--1) and $^{13}$CO(2--1) molecular gas \citep{Chou+07}. The associated molecular HCN emission is redshifted with respect to the systemic velocity of the galaxy, however, the velocity range of the emission, 570 -- 618\,km\,s$^{-1}$, falls short of the velocity covered by the class I methanol emission \citep{Cunningham&05}. \cite{Bendo+16} report free-free continuum and H$42\alpha$ images of NGC\,4945, however, this emission does not extend far enough south-east of the nucleus for the methanol maser to be associated.

The two regions of methanol maser emission in NGC\,253 have recently been suggested as being associated with the inner interface of the galactic bar and CMZ \citep{Ellingsen+submitted}. Like NGC\,253, NGC\,4945 is also classified as a barred spiral galaxy, however, in comparison to NGC253, the bar dynamics and molecular emission in NGC4945 are relatively poorly studied, due to a combination of a more edge-on orientation and a more southern declination. The assumed bar position angle is 33$^{\text{o}}$ with an azimuth angle with respect to the plane of the galaxy of 40$^{\text{o}}$ \citep{Ott+01}. Analysis of CO(2-1) maps indicate that the bar could come as close as 100\,pc to the galactic nucleus \citep{Chou+07}. 


Observations of hydrogen fluoride (HF) absorption towards NGC\,4945 by \citet{Monje+14} identify the possibility of molecular inflow towards the galactic centre. This HF inflow has a velocity range of 560 -- 720\,km\,s$^{-1}$ and \citeauthor{Monje+14} suggest an upper limit of $\sim200$\,pc on the radius of molecular inflow and maximum inflow velocity of 152\,km\,s$^{-1}$. The observed class I maser emission falls both within this velocity range and is located closer than the upper limit stated for the radius of inflow. The majority of redshifted molecular gas in NGC\,4945 is located north-east of the central region, therefore, the high redshift and south-eastern position of the class I emission could be explained through molecular inflow. In addition, this would be consistent with the explanation that class I methanol emission is associated with large-scale molecular in-fall, as observed in NGC\,253 \citep{Ellingsen+submitted}.

The only extragalactic source for which there has been a detailed investigation of the relationship between class~I methanol maser emission and other molecular lines is NGC\,253 \citep{Gorski+17, Ellingsen+submitted}. Hence, the reason why we observe 36.2\,GHz class I methanol maser emission offset to the south-east of the centre of NGC\,4945 and not at other locations is largely speculation.  Within the Milky Way, class II methanol masers are exclusively associated with high-mass star formation \citep{Breen+13b}, while class I masers (such as the 36.2\,GHz transition) are found predominantly in high-mass star formation regions, with a handful of sources detected towards low-mass star formation regions and supernova-molecular cloud interaction regions.  \citet{Breen+13b} present a detailed discussion of current knowledge on the mechanisms by which gas-phase methanol is created and destroyed.  Multiple lines of evidence suggest that methanol is formed on the surface of cold dust grains through hydrogenation of CO.  Observations of 36.2-GHz methanol masers in NGC\,253 and Arp~220 suggest that large-scale, low-velocity shocks are important, likely because they provide an efficient mechanism for releasing methanol from dust-grain mantles without dissociating it in the process.  Once released into the gas phase, methanol is depleted relatively quickly through a combination of gas-phase chemistry and/or dissociation from shocks or external radiation fields. We suggest that the combination of cold molecular gas and low-velocity shocks (both on large-scales) are the key requirements for the presence of a luminous class~I methanol maser emission, such as is observed in NGC\,4945.  Under this hypothesis, the absence of other sites of 36.2\,GHz methanol masers in NGC\,4945 is likely because other regions where there are low-velocity shocks do not also contain sufficient cold molecular gas.  In NGC\,253 the regions where the class~I methanol maser emission is observed show an enhancement in HNCO:SiO ratio \citep{Meier+15, Ellingsen+submitted}. This is thought to reflect the low-velocity shocks compared to more energetic shocks in molecular gas.  So we predict that observations of these two molecular tracers towards NGC\,4945 will likely show an enhancement of HNCO relative to SiO at the methanol maser location compared to the rest of the galaxy.

Although our sample size is not sufficiently large to perform any detailed analysis of trends that identify likely candidates for future searches of class I methanol emission, understanding the similarities between all known host galaxies may help future searches. This is the second barred-spiral starburst galaxy with class I emission potentially associated with molecular inflow. Therefore, we suggest targeting future searches for 36.2\,GHz emission in galaxies with similar properties.\\

\section{Conclusion} \label{sec:conc}

We present 36.2\,GHz ($4_{-1} \rightarrow 3_0 $E) imaging results of the starburst galaxy NGC\,4945. A region of 36.2\,GHz methanol emission is observed, offset from the galactic nucleus by 174\,pc, while emission from the 37.7\,GHz transition was not detected. Lack of strong observed thermal emission (from this same offset region) combined with narrow linewidth and high isotropic luminosity indicate that this region is a likely class I maser. The maser emission may be associated with molecular inflow,which is observed in HF absorption. If this is the case, it would fit with the explanations of similar emission in NGC\,253. This is the third detection of 36.2\,GHz masers in any extragalactic source, and the second within a barred starburst galaxy. As our sample size increases it will become easier to identify factors that make galaxies appropriate hosts for class I methanol masers, and allow higher success rates in targeted searches.

\acknowledgments
The Australia Telescope Compact Array is part of the Australia Telescope National Facility, which is funded by the Commonwealth of Australia for operation as a National Facility managed by CSIRO. This research has made use of NASA's Astrophysics Data System Abstract Service. This research also utilised APLPY, an open-source plotting package for PYTHON hosted at http://aplpy.github.com. This research made use of Astropy, a community-developed core Python package for Astronomy \citep{astropy+13}.

\bibliography{ref_ngc4945}

\end{document}